# Realizing Intelligence

Paul Yaworsky
*Information Directorate*
*Air Force Research Laboratory*
*Rome, New York 13441*

**Abstract**
Order exists in the world. The intelligence process enables us to realize that order, to some extent. We provide a high level description of intelligence using simple definitions, basic building blocks, a conceptual framework and general hierarchy. This perspective includes multiple levels of abstraction occurring in space and in time. The resulting model offers simple, useful ways to help realize the essence of intelligence.

*Keywords:* Intelligence, Cognition, Information, Abstraction, Realization, Framework, General Intelligence Hierarchy

# 1  An Orderly Process

Intelligence is an orderly process. The order stems from nature and exists everywhere, in space and in time. The order may be obvious or subtle or seemingly unknown. One of the main jobs of the intelligence process is to realize the order in nature, to some extent, thereby gaining a working knowledge of the world. This knowledge helps with survival, not to mention all the other aspects of human life. Yet our success at survival does not necessarily depend upon our individual level of intelligence. Nor do we need to understand how intelligence works to be intelligent. Intelligence exists in many different forms of life on this planet, varying from low to high levels, yet humans are believed to be the most intelligent life form, so we will focus here on high level (human) intelligence.

What do we really know about intelligence? What benefit is there from describing the process? How can the resulting knowledge improve our situation or help us accomplish our goals? Intelligence is like a big puzzle. There are very many pieces to the puzzle, they can be put together in so many different ways, and the pieces are dynamic (a function of time). While nature's order may be subtle, one thing the intelligence process does enable us to do is to put its pieces together to form an understanding of the world. However, putting the pieces together can lead to confusion, as complexity often overshadows underlying order, and the overall process involves complex relationships in space and time. There may be nothing in the intelligence process that says we need to understand the process itself. Yet we can try. Contained in nature's order are countless signals that feed the intelligence process. Over time these signals can reveal existing patterns, trends and tendencies. Humans have been able to learn much about many things over time, not the least of which is the underlying order contained in the intelligence process itself. We can also realize things at multiple levels of abstraction. The ability to abstract is one of the most significant aspects of intelligence, setting humans above all other animals in this respect.

Intelligence is a process that reflects nature's order. Admittedly the order is complex, but over time humans have developed the ability to unravel some of the complexities while learning more and more about the world. This leads to better realizations, which enable better descriptions, leading to better circumstances and greater opportunities… and the process continues. An overall job of the human brain is to exploit nature's order, to some extent, thereby forming a model of the external world (in the mind). At birth, the human cerebral cortex (primary region for high order intelligence) has a very large number of neurons loosely connected. Over time and through experience, the brain learns to represent nature's order internally, in its network of neurons, connections and layers. Neurons develop a propensity for firing (outputting) based on whatever activity is present. Connections form in a hierarchical, layered fashion. The higher up you go in the hierarchy, the richer the relationships and the higher the intelligence capabilities are possible. Overall, the intelligence process enables the brain to perceive sensed signals, learn patterns, represent activity, establish meaning and understand what is going on to some extent. Rich, dynamic relationships form as the brain organizes signals according to spatio-temporal patterns. These patterns and their sequences of activations form a model of the world often referred to as the mind.

## 2   Modeling Goals

Our approach to realizing intelligence involves developing a model of intelligence. By model we mean an approximation of the real thing. Real intelligence is tucked safely away within the skull. Our model will be an external, objective representation of the internal intelligence process. How well we describe the process depends on how well we understand and realize the nature of intelligence. We realize that we will fall far short of having a complete model. We also expect that over time many people will contribute and that much more will be learned about the nature of intelligence, in many different disciplines. For example, some key ideas described here leverage the work of Jeff Hawkins [1], with his goal aimed toward creating intelligent machines (computers) through an understanding of the brain. Improved understanding will lead to better descriptions and better implementations.

Our overall goal is to model intelligence using computers. This necessarily requires a sufficient understanding of intelligence. This also requires an understanding of computers, but since computers are designed and built by humans, we consider the computer portion of the goal to be understood well enough for now. Thus we do not focus on computers here. However, intelligence poses many problems. One of the underlying problems has to do with the nature of *information* and the confusing role it plays in the intelligence process. *Information* is a very common term with many different contexts and meanings, so we must be careful about how we use the term. We must also understand the differences between information processing performed by humans and computers, because the differences are fundamental and can have a significant impact on our viewpoints and resulting models. Finally, we need to develop computer technology that can assist humans in processing information more "intelligently." Ultimately this will involve top-down theoretical approaches, bottom-up detailed implementations and advanced information processing somewhere in-between.

## 3   Simple Descriptions

We have mentioned that intelligence is an orderly process, yet we acknowledge that the order is anything but simple. All too often the order is not obvious. The nature of intelligence has more or less eluded our grasp since the beginning of time (or at least since the beginning of intelligent life). But all is not lost. Humans have learned a lot about the intelligence process, and history reveals much. Generations of humans have passed down tremendous amounts of knowledge over the years, both explicitly and implicitly. Not only is the body of human knowledge significant, but our abilities to represent, store and communicate knowledge enrich the intelligence process. Built into the fabric of intelligence are the abilities to realize, understand, simplify and describe underlying processes. As such, we aim to do just that here (to the extent that we can), starting with some basic definitions. These definitions are given in the context of human intelligence.

> *Intelligence* is the ability to learn (to acquire knowledge, to abstract, to realize) and recall (apply what has been learned) effectively in a changing environment. To demonstrate intelligence, the learning must be applied or used effectively in a dynamic, real-world environment.
>
> *Cognition* basically means to know. As such, cognition is a subset of intelligence, and knowledge is a high-order construct of the overall process.
>
> *Abstraction* is the ability to capture the essence of something in a spatial and/or temporal sense. In a spatial sense, abstraction basically means to change form, for example as signal representations go from formal to formless (e.g., concept). In a temporal sense, abstraction means to change state, for example as signal representations go from timely to timeless. Abstraction represents a significant and orderly kind of change, and leads to powerful forms of learning.

*Information* is an intermediate kind of signal representing relative spatial or temporal relationships. This makes information a mid-level construct of the intelligence process. Other definitions and contexts exist for information, for example those involving computers and communication systems.

*Decision* is a choice or conclusion. Basically a decision is the act of considering many options and selecting one (or some function of them). This definition stems from the very nature of neurons in the brain. Neurons have many inputs and one output. This "many-to-one" function is fundamental to intelligence, occurring everywhere from low to high levels of brain activity. Neurons make decisions for a living.

*Meaning* refers to signal relationships or associations in the brain. The brain accomplishes meaning by way of connectivity along with associated neural activity. Connectivity has a direct impact on meaning since it literally forms relationships among signals. However, since the brain has billions of neurons and trillions of connections, the number of possible combinations of activity is enormous. This helps explain why meaning can and usually does differ among people, since no two people have the exact same mental model or the same set of signals activated at the same time.

*Representation* is a form of description, designation or memory of signal activity occurring in the brain. Representation not only supports real-time processing but also enables various kinds of storage and communication of signals. Once signals have been *presented* to the brain, they can then become *re-presented* by the mind. Spatio-temporal patterns can be learned, represented and combined to form sophisticated models of the world.

Our goal is to model intelligence using computers. We describe our model in a top-down manner using fundamental concepts. We also use basic building blocks in a bottom-up fashion. While these descriptions are simple, they serve as basic components in the construction of our model. Furthermore, the choice of which pieces to use and how to put them together can have a tremendous impact on the resulting model. In the end, we aim for descriptions that provide an overall picture of intelligence and capture the essence of the process. These descriptions will provide the foundation for a better understanding of the intelligence process and lead to the development of better computational models.

# 4  Basic Building Blocks

We now describe a set of basic building blocks for our model, inspired by the brain. The first building block is the *signal* – a basic component of intelligence. In neuroscience, the term *signal* is often referred to as a spike, pulse, impulse or action potential. However, we use the term *signal* in a more generic way, to describe any kind of communication or transfer of activity within our model. This includes external signals (inputs to and outputs from the model) as well as internal signals (transfer of activity within the model). Information is another term often used to describe *signal* activity, but we view information in a narrower context. We will describe later what we mean by this, but basically information refers to intermediate kinds of signals that occur somewhere between lower and higher order signals within the intelligence process.

The *neuron* is the main processing unit in the brain. The neuron is a type of living cell. Neurons typically consist of a cell body, many dendrites to handle the inputs and an axon to handle the output. This basic structure can vary in different regions of the brain. As signals enter the neuron, each input is processed along with any prior activity stored in the neuron. The neuron's job is basically to process many inputs and produce a single output at any given time. Thus each neuron basically performs a many-to-one function by virtue of its many-to-one structure. The output signal is a composite, abstract representation of the activity entering and occurring within the neuron at any given time.

Another fundamental building block in the brain is the *connection*. A connection literally connects one neuron to another. However, these connections (especially within the cerebral cortex) usually are not solid, hard-wired passage ways. Instead, neural connections typically contain a gap or synapse. The synapse does its own processing, thereby serving as a memory unit by virtue of electro-chemical activity occurring across the gap. Each synapse can also act like a switch, since a signal will either pass or not pass, depending on the signal's energy level along with any preset conditions at the synapse.

*Hierarchy* is another building block in our model. Hierarchies are cumulative in nature, formed by combinations of neurons and connections. In the brain, many hierarchies exist by virtue of many components being arranged and activated in an orderly fashion. Multiple levels may exist within each hierarchy. At lower levels, many neural entities handle specific details at mostly local scales. At higher levels, fewer yet more significant entities exist within the model. In a larger sense, hierarchies represent a scaled up version of the many-to-one processing that occurs in a single neuron. As signals proceed up the ranks, hierarchies come to represent a natural kind of order (i.e., patterns) in terms of brain function as well as structure. As such, hierarchies are fundamental to intelligence.

# 5  Framework

Now that we have described some basic components and simple definitions for our model, we can put them together to form a basic framework (see Figure 1). Although conceptual in nature, this framework does provide a high level view of what the intelligence process may look like. The framework is oriented with two coordinates or dimensions, one that corresponds with signals changing in a spatial sense and the other with signals changing in a temporal sense. Within each dimension, signals become organized into relative hierarchies. Within each hierarchy, a many-to-one reduction occurs as signals proceed upward, going from lower to higher levels. By definition, higher level entities are more significant than lower level ones. Many input signals originate in the external, physical world. Or, so-called input signals can be the result of prior internal processing, occurring in the form of feedback. Of course, many combinations of signals can occur. All signals get processed and transformed as they flow upward within their respective hierarchies. By virtue of hierarchies, this bottom-up activity is another example of many-to-one processing that occurs all across the brain. Output signals, on the other hand, generally flow downward within a hierarchy. These downward signals may go out to the external, physical world (as an external application or use of intelligence), or they may be used internally in the form of feedback to influence bottom-up learning, or both.

In the spatial dimension of this framework, signals have a physical presence and physical characteristics. From an external point of view, signals originating in the physical world have physical form (e.g., formal, physical, concrete). This includes all aspects of three spatial dimensions. However, once signals enter the model in a spatial sense, they become transformed into more generic-like signal representations. The activations and combinations of these signals form patterns, and the patterns can be learned. As lower level signals combine and travel up their respective hierarchies to form higher level signals (realizations), they continue to be processed and transformed. Along the way, the physical *form* and initial spatial attributes of signals get abstracted out, and ultimately all that remains is their *essence*. At high levels, this results in things such as concepts, beliefs and theories. These things represent the essence of signals caused by, or coming from, the external world. In essence these high level signals have no form. This transformation process (i.e., change), whereby signals go from representing things with physical form to representing things with no form, is called spatial abstraction. Relative to the external world, physical form gets abstracted out of signals. Therefore, resulting high level signals become *form*-invariant.

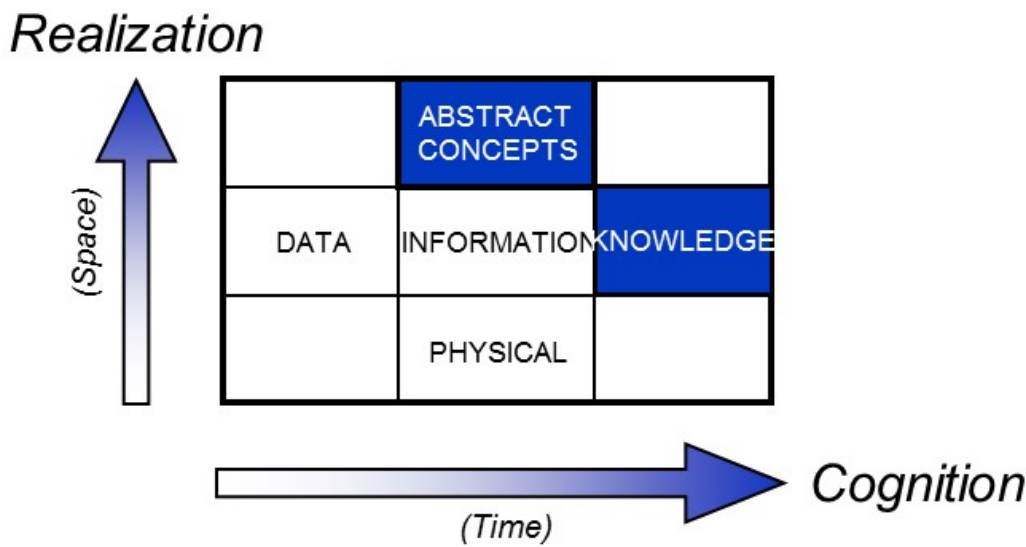

**Figure 1.** Framework for modeling intelligence in space and time.

In the temporal dimension, signals take on relative significance with respect to time. As nature would have it, signals in the external, physical world have characteristics that are timely, time sensitive and/or quickly changing, relatively speaking. Such temporal characteristics can also be thought of in terms of frequency (how often a signal may change value) or usefulness (how long a signal's value may be useful). As signals flow upward within temporal hierarchies, they transition into higher and higher level states (e.g., signals become more and more timeless, slow to change or time-insensitive). Higher level signals come to represent the essence of signals with respect to time. This process, whereby signals transition (i.e., change) from representing time-sensitive things to time-insensitive things, is a form of temporal abstraction. As signals transition up their respective hierarchies, *time* gets abstracted out, and ultimately all that remains is their *essence*. Therefore, resulting high level signals become *time*-invariant. For example, the *essence* of signals in a temporal sense may include things such as general knowledge, basic principles and fundamental theories. Signals also take on temporal significance by virtue of sequences. A sequence is an orderly series of signals occurring over time. Within a temporal hierarchy, lower level signals may represent snapshots in time or perhaps shorter sequences. Higher level signals represent larger timeframes or longer sequences. These higher level constructs naturally represent more of the input space and thus have more temporal significance. The intelligence process is fundamentally a function of time, and temporal characteristics of signals are essential to the process.

Intelligence hinges upon knowledge, and since cognition basically means "to know," cognition has a prominent place in our framework. That is, the intelligence process leads toward the formation and use of knowledge over time. We view this as a form of temporal abstraction. In a spatial sense, the realization of concepts is a form of spatial abstraction, also fundamental to intelligence. Thus abstraction (i.e., change) occurs with respect to both space and time, and hierarchies form accordingly. As part of the hierarchical processes, many basic signals exist at lower levels, and fewer, more significant signals exist at higher levels. In a spatial sense, high level realizations such as concepts, key points, big ideas and theories are essential to human intelligence. In a temporal sense, high level constructs such as knowledge, theories and long term perspectives are essential. Of course, actual brain signals are inextricably coupled in space and time, and not independent in either context. This framework is merely an artifact, a rough model used to help understand and describe the intelligence process. This model will be modified over time as new knowledge of brain structure develops and as we gain a better understanding of brain function.

# 6 Basic Levels of Realization

Our conceptual framework reveals general properties about the intelligence process that become obvious at a high level. From an overall perspective, three basic levels of realization become evident. Whether considered from a spatial or temporal point of view, the highest level of realization (Level 1) is the most "abstract." That is, the abstract level deals with the essence of things (e.g., concepts, knowledge) as opposed to the formal "here and now" nature of signals in the physical world. At the most abstract level, spatial characteristics are transformed (changed) into signals which have no physical form, and temporal characteristics are transitioned (changed) into signals which are timeless (relatively speaking). However, to the extent that we can describe these abstractions (or describe anything for that matter), a formal language or system of external representations is needed to "inform" the process. These kinds of representations occur at a middle level (Level 2), involving realizations often referred to as "information." While information is a generic, ubiquitous term, it does serve as a useful descriptor, thereby having a key role within our framework. Resulting descriptions may be tailored and applied to any level of abstraction. Finally, the lowest level of realization (Level 3), in both a spatial and temporal context, is "physical." This level provides the physical basis for internal signal activity as well as the interface for interaction (Input/Output) with the external, physical world. As such, the physical level supports all levels of signal activity and realization.

# 7 Intelligence Hierarchy

The three basic levels of realization mentioned above, as well as the spatial and temporal hierarchies in our conceptual framework, can be combined into a single, general hierarchy of intelligence (see Figure 2). This overall hierarchy helps us to further understand and describe the intelligence process. The lowest level in the hierarchy represents physical signal activity, and is therefore most closely associated with the external world. The physical level also enables higher levels of realization within our model. Built upon the physical level is the so-called information level. Information is a relative, generic term that can be used to describe signal activity internally (connectivity) and externally (communication). Language is an important example of this kind of signal activity. At an even higher level within the intelligence hierarchy are abstract realizations. These abstractions are built upon the physical and information levels of realization and are the result of all kinds of internal activity. As part of human intelligence, these three basic levels of activity are not independent, but must work together. That is, physical signals are used to implement and enable all three levels of realization; information is used to describe activity at all three levels; and abstractions occur as signals are transformed across all three levels of realization. Our ability to understand is a function of the interaction and harmony of signal activity occurring among levels (i.e., top-down/bottom-up resonance). Of course, a tremendous amount of activity can occur at any level, and many sub-levels and hierarchies can also exist.

Within this simple hierarchy, order is implicit, with increasing intelligence capabilities established as you go upward. By definition, higher level entities (and their activations) are more significant than lower level ones. Many input signals enter the hierarchy from the external, physical world. Or, so-called input signals may be activated by way of prior internal processing, occurring in the form of feedback. Either way, this can result in much internal signal activity, with many combinations of signals possible. Input signals generally flow upward within the hierarchy, and this bottom-up activity is called learning. Overall, the learning process transforms (changes) very many signals into fewer, more abstract realizations. On the other hand, output signals generally flow downward within the hierarchy. These downward signals may go out to the external, physical world, or they may be used internally in the form of feedback to influence bottom-up learning, or both. Output signals represent the application or use of that which has previously been learned (this is the second part of our definition of intelligence). Overall signal flow is shown in Figure 2. The resulting complexity is literally mind-boggling (and we are not even talking here about other-than-conscious activity).

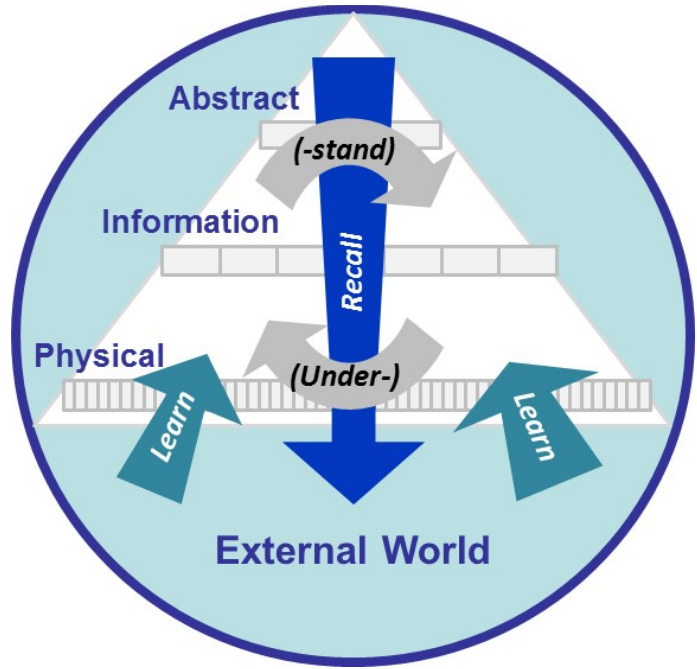

**Figure 2.** General intelligence hierarchy.

The following set of examples can help illustrate the development of intelligence as an orderly progression involving three levels of realization. At the lowest (physical) level is matter and energy. Of course, matter and energy exist in many forms and have many inherent temporal properties, serving as fundamental building blocks in nature. At the middle level in our example is life. Life-forms are a special kind of matter and energy consisting of cells organized in such ways as to enable life-processes. Each living cell contains information coded in the form of DNA, RNA, etc. used to *inform* matter and energy what to do in order to sustain life. While individual cells work on small scales, groups of cells are organized and combined to form more complex, dynamic, hierarchical systems. This can result in many kinds of life-forms, with varying capabilities. Thus life is enabled by virtue of interactions between the physical and information levels of realization. However, at an even higher level and even more significant is a special kind of life – intelligent life. Intelligence requires the ability to perform abstract realizations (Level 1) of information (Level 2) based on underlying physical activity (Level 3). Neurons are special kinds of cells that enable these high level abstractions. Neurons and their connections are configured in a hierarchical fashion, thereby providing the ability to realize higher levels of abstraction (e.g., knowledge, concepts) as well as the ability to recall (apply or use) those abstractions effectively in a changing environment. Thus neurons perform abstractions at all levels. Of course, higher level intelligence requires lower level capabilities which are enabled by all levels of realization. This makes intelligence a coordinated, cumulative, hierarchical process based upon many capabilities.

These three levels of realization and abstraction are somewhat analogous to the "Three Levels" described by David Marr [2]. While Marr was interested mainly in modeling vision using computational methods, his "Three Levels" approach did indicate the need to understand information processing tasks at various levels of analysis and description. Furthermore, he indicated that if an "information processing" task was not understood well enough at Level 1, then computational modeling efforts focused at lower levels may not work well enough. We realize the significance of Marr's Level 1 in our model, acknowledging the need for top-down as well as bottom-up approaches in our efforts. After Marr, another way to state the three levels needed to model intelligence are simply: realize, describe and implement.

We close by reemphasizing our specific use of the term *information* in the general intelligence hierarchy. Within the spectrum of processing, information is an intermediate, relative term. Within our model, information exists between lower (physical) and higher (abstract) levels of processing. While signals can and do exist internally in various forms and at different levels, the only way that humans can process and describe the corresponding activity *externally*, whether signals originate in the physical world or in the abstract realm, is by using information. This makes information a special kind of signal that can be recognized, processed, acted upon, stored and communicated. Ultimately, information is associated with all kinds of descriptions, representations and languages. Yet we should not forget that the overall purpose of "information" is to support intelligence. Within the whole scheme of things, information is a "means to an end." The end result is intelligence.

# 8 Conclusion

The overall goal is to model intelligence using computers. However, computer technology is not the limiting factor here. What's needed is a sufficient understanding of the intelligence process itself. We must be able to realize fundamental concepts, describe the overall process as clearly as possible, and implement the process in a physical system. Top-down realizations must be coupled with bottom-up implementations. Obviously, we have a long way to go in our modeling efforts. But the description provided here does offer a simple, useful perspective of intelligence, albeit at a high level.

Intelligence is a process whose order is derived from nature. Signals are processed by the brain in an orderly fashion, undergoing sophisticated abstractions (changes) in space and in time. Resulting signals are organized hierarchically, with many basic components at lower levels giving way to fewer yet more significant entities higher up. From a steady stream of input signals, many lower level pieces combine to form higher level pieces. Each piece can contribute to the harmony of signal patterns occurring in the brain. We describe a framework that provides a big picture of the overall intelligence process in terms of space and time. We discuss three levels of realization that represent the kinds of processing occurring in intelligence. Finally, we simplify these representations into a general intelligence hierarchy that incorporates relevant underlying properties. Within this general hierarchy, the lowest level of realization is physical, the intermediate level is called information, and the highest level is the most abstract. While each level is limited in what it represents, all levels must work together to enable the realization of intelligence.